\documentclass[a4paper,12pt]{article}
\usepackage[english]{babel}
\usepackage[latin1]{inputenc}

\usepackage{cite}
\usepackage{amsmath,amssymb,amsfonts}
\usepackage{latexsym}

\newif\ifpdf
\pdffalse 

\usepackage{graphicx}

\newcommand{\be}{\begin{equation}}
\newcommand{\ee}{\end{equation}}
\newcommand{\e}{\boldsymbol{e}}
\newcommand{\bx}{\boldsymbol{x}}
\newcommand{\bk}{\boldsymbol{k}}
\newcommand{\phat}{\widehat{p}}
\newcommand{\qhat}{\widehat{q}}
\newcommand{\khat}{\widehat{k}}

\newcommand{\scrd}{{\cal D}}
\newcommand{\scrh}{{\cal H}}

\def\psibar{{\bar{\psi}}}
\def\bsigma{{\boldsymbol{\sigma}}}

\def\bpi{{\boldsymbol{\pi}}}

\def\bx{{\boldsymbol{x}}}
\def\bp{{\boldsymbol{p}}}
\def\bq{{\boldsymbol{q}}}
\def\smfrac#1#2{{\textstyle{\frac{#1}{#2}}}}

\newcommand{\<}{\langle}
\renewcommand{\>}{\rangle}

\newcommand{\reff}[1]{(\ref{#1})}

\title{Renormalization flow for unrooted forests on a triangular lattice}
\author{ Sergio Caracciolo\\
\small\it  Universit\`a degli Studi di Milano - Dip.~di Fisica and INFN,\\ [-0.2cm]
\small\it  via Celoria 16, I-20133 Milano, Italy\\ [-0.2cm]
\small\tt Sergio.Caracciolo@mi.infn.it      \\ [-0.2cm]
{\protect\makebox[5in]{\quad}}
\and 
 Claudia De Grandi\\ [-0.2cm]
 \small\it Departement of Physics, Boston University, Boston, MA 02215, USA
 \\ [-0.2cm]
\small\tt degrandi@bu.edu      \\ [-0.2cm]
{\protect\makebox[5in]{\quad}}
\and
Andrea Sportiello\\
\small\it  Universit\`a degli Studi di Milano - Dip.~di Fisica and INFN,\\ [-0.2cm]
\small\it  via Celoria 16, I-20133 Milano, Italy \\ [-0.2cm]
\small\tt Andrea.Sportiello@mi.infn.it    
}
\begin{document}

\ifpdf
\DeclareGraphicsExtensions{.pdf, .jpg, .tif}
\else
\DeclareGraphicsExtensions{.eps, .jpg}
\fi

\maketitle

\thispagestyle{empty}   

\begin{abstract}
We compute in small temperature expansion the two-loop renormalization
constants and the three-loop coefficient of the $\beta$-function, that
is the first non-universal term, for the $\sigma$-model with $O(N)$
invariance on the triangular lattice at $N=-1$.  The partition
function of the corresponding Grassmann theory is, for negative
temperature, the generating function of unrooted forests on such a
lattice, where the temperature acts as a chemical potential for the
number of trees in the forest.  To evaluate Feynman diagrams we extend
the coordinate space method to the triangular lattice.
\end{abstract}


\clearpage
\section{Introduction}

Results concerning with graph 
theory~\cite{Temperley, Biggs, Royle, Diestel}, that is properties of
a set of points which refer simply to
the notion of adjacency, are of interest in a variety of fields,
ranging from pure mathematics to statistical physics and find an
enormous amount of applications in natural sciences besides physics
like in biology or in theoretical information science.

Detailed properties of a graph can be derived from the study of the
partition function of a $q$-state Potts model~\cite{Potts,Wu,Wu2} with
variables defined on its sites. Indeed this function is strictly
related with the Tutte polynomial of the
graph~\cite{tutte,foata,alantutte} and, for example, the generating
polynomial of spanning trees or unrooted forests on the graph can be
recovered by taking the limit $q\to 0$.

A classical result in algebraic graph theory is Kirchhoff's
matrix-tree theorem~\cite{Kirchhoff} which expresses the generating
polynomials of spanning trees and rooted spanning forests on a given
graph as determinants associated to the graph's Laplacian matrix. For
recent applications see for example~\cite{Schrock,Glasser}.  It is
quite natural to rewrite these determinants as Gaussian integrals over
Grassmann variables.

Recently~\cite{noi} it has been shown that the solution of other
combinatorial problems on a graph can be represented in terms of
Grassmann integrals, eventhough non-Gaussian. In particular, the
generating polynomial of unrooted spanning forests on the graph is
simply written adding to a Gaussian term a suitable four-fermion
term. Interestingly, the same partition function can be obtained,
order by order in perturbation theory, by considering an
anti-ferromagnetic non-linear $\sigma$-model with $O(N)$ invariance in
the limit in which $N\to -1$. These representations are very
convenient to study the cases in which the graph is an infinite
regular lattice, because the whole machinery of Statistical Field
Theory becomes avalaible. For example, Renormalized Perturbation
Expansion can be used, Renormalization Group notions can be applied
and one sees that on two dimensional lattices these models are
asymptotically free~\cite{Polyakov_75, brezin, Brezin_76, Bardeen_76}.
The same mapping has been used at the transition at negative tree fugacity which corresponds to the Potts antiferromagnetic critical point~\cite{saleur1,saleur2,saleur3}.

In this paper we will concentrate on the triangular lattice and, in
particular, we are interested in the evaluation of the so-called
$\beta$-function. We have computed the three-loop coefficient which is
the first non-universal term, which, in contrast with the square
lattice, was yet unknown. A direct practical relevance of this
coefficient comes from a recent study of the zeroes in the complex
plane of the partition function of the Potts model by means of the
numerical evaluation of a transfer matrix in a
strip~\cite{alanpotts}. The locum of zeroes converges to a pair of
complex-conjugate curves with horizontal asymptote, but the
convergence is very slow in a region of large 
${\rm Re}(w)$.
It turns out that the shape of this curve
can be deduced perturbatively (in $1/w$) from the expression of the
$\beta$-function, thus in the region where the errors are larger.

\section{Unrooted forests}
\label{uforests}

Let $G=(V,E)$ be a finite undirected graph with vertex set $V$ and edge set $E$.  
Associate to each edge $e$ a weight $w_e$, which can be a real or complex number or, more generally,
a formal algebraic variable.
For $i \neq j$, let $w_{ij} = w_{ji}$ be the sum of $w_e$ over all edges $e$ that connect $i$ to $j$.
The (weighted) Laplacian matrix $L$ for the graph $G$
is then defined by 
\be
L_{ij} = \begin{cases} 
-w_{ij} & \text{for } i \neq j,\\
 \sum_{k \neq i} w_{ik} & \text{for } i = j \, .
 \end{cases}
 \ee
This is a symmetric matrix with all row and column sums equal to zero.

Since $L$ annihilates the vector with all entries 1,
its determinant is zero.
Kirchhoff's matrix-tree theorem \cite{Kirchhoff}
and its generalizations \cite{Chaiken,Moon,Abdesselam,hyperforests}
express determinants of square submatrices of $L$
as generating polynomials of spanning trees
or rooted spanning forests in $G$.
For any set of vertices $\{i_1,\ldots,i_r\}$ of $V$,
let $L(i_1,\ldots,i_r)$ be the matrix obtained from $L$ by deleting
the  rows and columns $i_1,\ldots,i_r$.
Then Kirchhoff's theorem states that $\det L(i)$ is independent of $i$
and equals
\be
   \det L(i)  \;=\; \sum_{T \in {\cal T}} \, \prod_{e \in T} w_e   \;,
 \label{eq.Kirchhoff}
\ee
where the sum runs over all spanning trees $T$ in $G$.
(We recall that a subgraph of $G$ is called a tree if it
 is connected and contains no cycles,
 and is called spanning if its vertex set is exactly $V$.)
The $i$-independence of $\det L(i)$ expresses,
in electrical-circuit language,
that it is physically irrelevant which vertex $i$
is chosen to be ``ground''.
There are many different proofs of Kirchhoff's formula \reff{eq.Kirchhoff};
one simple proof is based on the Cauchy--Binet theorem
in matrix theory (see e.g.\ \cite{Biggs}).

The ``principal-minors matrix-tree theorem'' reads
\be
   \det L(i_1,\ldots,i_r)  \;=\;
   \sum_{F \in {\cal F}(i_1,\ldots,i_r)} \, \prod_{e \in F} w_e   \;,
 \label{eq.principal}
\ee
where the sum runs over all spanning forests $F$ in $G$
composed of $r$ disjoint trees, each of which contains exactly one
of the ``root'' vertices $i_1,\ldots,i_r$.
This theorem can easily be derived by applying
Kirchhoff's theorem \reff{eq.Kirchhoff} to the
graph in which the vertices $i_1,\ldots,i_r$
are contracted to a single vertex, while it has theorem
(\ref{eq.Kirchhoff}) as a special case $r=1$, through the bijection
between unrooted spanning trees and spanning trees rooted on a given
fixed vertex.

Let us now introduce, at each vertex $i \in V$,
a pair of Grassmann variables $\psi_i$, $\psibar_i$.
All of these variables are nilpotent ($\psi_i^2 = \psibar_i^2 = 0$),
anticommute, and obey the usual rules for Grassmann integration.
Writing 
\be
\scrd(\psi,\psibar) := \prod_{i \in V} d\psi_i \, d\psibar_i \, ,
\ee
we have, for any matrix $A$,
\be
   \int \scrd(\psi,\psibar) \; e^{\psibar A \psi}
   \;=\;
   \det A
\ee
and more generally
\be
        \int \! \scrd(\psi,\psibar) \;  
        \bigg( \prod_{\alpha=1}^r
	\psibar_{i_\alpha} \psi_{i_\alpha}
               \bigg)
                \, e^{\psibar A \psi}
 =\,   \det A(i_1,\ldots,i_r)
   \; .
\ee
These formulae allow us to rewrite the matrix-tree theorems in Grassmann form;
for instance, \reff{eq.Kirchhoff} becomes
\be
   \int \! \scrd(\psi,\psibar)
            \,  \psibar_{i} \psi_{i}
               \, e^{\psibar L \psi}
   \;=
  \sum_{T \in {\cal T}} \, \prod_{e \in T} w_e  \;.
 \label{eq.principal.2}
\ee
while \reff{eq.principal} becomes
\be
   \int \! \scrd(\psi,\psibar) \,
               \bigg( \prod_{\alpha=1}^r \psibar_{i_\alpha} \psi_{i_\alpha}
               \bigg)
               \, e^{\psibar L \psi}
   \;=
   \sum_{F \in {\cal F}(i_1,\ldots,i_r)} \, \prod_{e \in F} w_e 
    \label{eq.principal.3}
\ee
which is to say 
\be
     \int \! \scrd(\psi,\psibar)
        \, \exp \Big[
          \psibar L \psi \,
         +\, t \sum\limits_i \psibar_i \psi_i\,
                       \Big]
   = 
       \sum_{\substack{
                F \in {\cal F} \\
                F = (F_1,\ldots,F_{\ell})
          }}
      \, t^{\ell}\,
       \bigg( \prod_{i=1}^{\ell} \,  |V_{F_i}| \bigg)
       \, \prod_{e \in F} w_e  \;.
 \label{eq.t}
\ee
This formula represents vertex-weighted spanning forests
as a massive fermionic free field \cite{Duplantier_88,Biggs}.

More generally, it has been shown in~\cite{noi} that 
\begin{multline}
     \int \! \scrd(\psi,\psibar)
        \, \exp\!\Big[
          \psibar L \psi \,
         +\, t \sum\limits_i \psibar_i \psi_i\,
         +\, u \sum\limits_{\< ij \>} w_{ij} \psibar_i \psi_i \psibar_j \psi_j
               \Big]
\\
   = \!\!\!\!\!\!
       \sum_{\substack{
                F \in {\cal F} \\
                F = (F_1,\ldots,F_{\ell})
          }}
       \!\!\!\!\!
       \bigg( \prod_{i=1}^{\ell} \,  (t|V_{F_i}| + u|E_{F_i}|) \!\bigg)
       \,
       \prod_{e \in F} w_e
 \label{eq.fourfermion}
\end{multline}
where the sum runs over spanning forests $F$ in $G$
with components $F_1,\ldots,F_{\ell}$;
here $|V_{F_i}|$ and $|E_{F_i}|$ are, respectively,
the numbers of vertices and edges in the tree $F_i$.
We remark that the four-fermion term
 $u \sum_{\< ij \>} w_{ij} \psibar_i \psi_i \psibar_j \psi_j$
 can equivalently be written, using nilpotency of the Grassmann variables,
 as $-(u/2) \sum_{i,j} \psibar_i \psi_i L_{ij} \psibar_j \psi_j$.
More interestingly,
since $|V_{F_i}| - |E_{F_i}| = 1$ for each tree $F_i$,
we can take $u=-t$
and obtain the generating function of {\em unrooted}\/ spanning forests
with a weight $t$ for each component.

\section{Relation with the lattice $\sigma$-Models.}

Recall that the $N$-vector model consists of spins $\bsigma_i \in \mathbb{R}^N$,
$|\bsigma_i| = 1$, located at the sites $i \in V$,
with Boltzmann weight $e^{-\scrh}$ where
\be
\scrh = -T^{-1} \sum_{\<ij\>} w_{ij} (\bsigma_i \cdot \bsigma_j - 1) \label{scrh}
\ee 
and $T$ is the temperature.

Low-temperature perturbation theory is obtained by writing
\be
\bsigma_i = (\sqrt{1-T\bpi_i^2}, T^{1/2} \bpi_i)
\ee
with $\bpi_i \in \mathbb{R}^{N-1}$ and expanding in powers of $\bpi$.
Taking into account the Jacobian,
the Boltzmann weight is $e^{-\scrh'}$ where
\begin{eqnarray}
   \scrh' & = & \scrh \,+\, \frac{1}{2} \sum\limits_i \log(1-T\bpi_i^2)
        \\
   & = &
   \frac{1}{2} \sum\limits_{i,j} L_{ij} \bpi_i \cdot \bpi_j
   \,-\, \frac{T}{2} \sum\limits_i \bpi_i^2
   \,-\, \frac{T}{4} \sum\limits_{\<ij\>} w_{ij} \bpi_i^2 \bpi_j^2
   +\,  O(\bpi_i^4, \bpi_j^4)
   \;.
\end{eqnarray}
When $N=-1$, the bosonic field $\bpi$ has $-2$ components,
and so, at least in perturbation theory, it can be replaced by a
fermion pair $\psi,\psibar$ if we make the substitution
\be
\bpi_i \cdot \bpi_j \to \psi_i \psibar_j - \psibar_i \psi_j \, .
\ee
Higher powers of $\bpi_i^2$ vanish due to the nilpotence of the
Grassmann fields,
and we obtain the model \reff{eq.fourfermion}
if we identify 
\be
t \,=\, -u \,=\, -T  \, .
\ee
Note the reversed sign of the coupling:
the spanning-forest model with positive weights ($t > 0$)
corresponds to the {\em anti}\/ferromagnetic $N$-vector model ($T < 0$).

In the case of a {\em regular} unweighted graph of order $q$, that is
all the vertices are connected to other $q$ vertices, we shall take 
\be
w_{ij} = \begin{cases} 1 & \hbox{ if $i$ and $j$ are connected}\\
0 & \hbox{ otherwise.}
\end{cases}
\ee
and the corresponding Laplacian 
\be
L_{ij} = 
\begin{cases} 
-1 & \hbox{ if $i\neq j$ and $i$ and $j$ are connected}\\
0 & \hbox{ if $i\neq j$ and $i$ and $j$ are not connected}\\
q & \hbox{ if $i = j$}\, .
\end{cases}
\ee
This is the case of a regular periodic lattice in $d$ dimensions. If
we take unit lattice spacing, vertices connected to a given site
correspond to sites at unit distance, 
so that, if $\hat{e}_k$ is a lattice direction, $f$ a lattice function,
and $x$ a lattice site, the lattice derivatives are defined as
\begin{align}
\nabla_{k} \, f(x) &:=  f(x+\hat{e}_{k}) - f(x)\\
\nabla^{*}_{k} \, f(x) &:=  f(x) - f(x-\hat{e}_{k}) 
\end{align}
The Laplacian can be written as
\be
(L f)(x) := - \sum_{k=1}^{q} \nabla_k f(x) 
\ee
and when, like in the square and triangular lattice, $q$ is even and to each lattice direction corresponds an inverse lattice direction, that is $\hat{e}_{k+q/2}= -\hat{e}_k$, we can restrict the sum to {\em positive directions}
\begin{align}
-(L f)(x) = & \sum_{k=1}^{q/2} \left(\nabla_k -\nabla_k^*\right) f(x) =  \sum_{k=1}^{q/2} \nabla_k \nabla_k^* f(x)  =  \sum_{k=1}^{q/2} \nabla_k^* \nabla_k f(x) \\
= &  \sum_{k=1}^{q/2} \left[ f(x+\hat{e}_k) + f(x-\hat{e}_k) - 2 f(x) \right]
\end{align}
and, in the lattice scalar product
\be
(f,g) = \sum_x f(x) g(x)
\ee
we have
\be
(g, L f) = - \sum_{k=1}^{q/2}  (g, \nabla_k^* \nabla_k f) = \sum_{k=1}^{q/2} (\nabla_k g,  \nabla_k f) = \sum_{k=1}^{q/2} (\nabla_k^* g,  \nabla_k^* f)\, .
\ee

\section{The calculus of the $\beta$-function}
\label{betacalc}

We follow a procedure which already found several applications
\cite{falctreves,3loop,4loop,shin2,4-loop_bis} for the square lattice.
For a lattice theory, i.e. a theory regularized by introducing a
discretization of the coordinates space, in principle the
$\beta$-function can be found by a direct computation on the lattice,
which also provides a regularization. However, our lattice
$\sigma$-model has a natural continuum counterpart, with the widely
investigated action
\be
\label{model}
\mathcal{S}(\bpi,h)=\beta \int d^2x \left[
\frac{1}{2}(\partial_{\mu}\bpi(x))^2+\frac{1}{2}
\frac{ \big( \bpi(x) \cdot \partial_{\mu}\bpi(x) \big)^2}
{1-\bpi^2(x)}-h\sqrt{1-\bpi^2 (x)}\right] \;, 
\ee
where we have introduced an external magnetic field $h$ which
explicitly breaks the $O(N)$-invariance. In particular Br\'ezin and
Hikami~\cite{brezin3loop} already performed the renormalization up to
three loops in dimensional regularization.

A general theorem of Renormalization states that the $n$-loop
$\beta$-function within a certain regularization scheme can be deduced
from the knowledge of the $\beta$-function in any other scheme, at the
same perturbative order, and of the renormalization constants in the
desired scheme, up to order $n-1$. So, a possible procedure, which we
will indeed follow in this work, is to relate the $\beta$-function on
the square and triangular lattice to the continuum results of Br\'ezin
and Hikami via the calculation of the two renormalization constants of
the non-linear $\sigma$-model, denoted by $Z_{1}$ and $Z_{2}$.

More in detail, in our case we have to compare our lattice theory with
the continuum theory renormalized in \cite{brezin3loop} using
$\overline{MS}$-scheme (Minimal Subtraction modified) and in
dimensional regularization.  The starting point is the relation for
the $n$-point 1-particle-irreducible (1PI) correlation functions
\be
\label{rino}
\Gamma_{latt}^{(n)}
\big( p_{1},\cdots ,p_{n}; \beta, h; 1/a \big)
=
Z_{2}^{n/2}
\, \Gamma_{\overline{MS}}^{(n)}
\big( p_{1},\cdots, p_{n}; Z_{1}^{-1}\beta, Z_{1}Z_{2}^{-1/2}h; \mu \big)
\ee
where $a$ and $\mu$ are respectively the lattice spacing and the scale
of renormalization for the continuum, while $p_{1},\ldots , p_{n}$ are
the external momenta.  Here we consider the lattice theory (denoted by
subscript {\it latt}) as a regularization of the continuum theory
renormalized at the scale $1/a$ and we compare it with the continuum
theory renormalized in the ${\overline{MS}}$-scheme (denoted by
subscript $\overline{MS}$) at the scale $\mu$ to determine the finite
constants $Z_{1}(\beta,\mu a)$ and $Z_{2}(\beta,\mu a)$.
%
Both the regularized theories satisfy a Renormalization Group equation:
\begin{align}
\frac{ \mathrm{d} }{ \mathrm{d} \mu}\Gamma_{\overline{MS}}^{(n)}
&=0 \,;
&
-\frac{ \mathrm{d} }{ \mathrm{d} a}\Gamma_{latt}^{(n)}
&=0
\,;
\end{align}
where we added a minus sign for the lattice equation, because when $a
\to 0$ we are making a RG flux toward short distances behaviour, that
has the reversed sign respect to the $\mu \to \infty$ limit made for
the continuum theory. 
%
For the lattice theory
\begin{multline}
0= -a \frac{ \mathrm{d} }{ \mathrm{d} a}\Gamma_{latt}^{(n)}
= \bigg[ -a 
\frac{\partial}{\partial a}
+W^{latt}(\beta)\frac{\partial}{\partial \beta^{-1}}
\\
- \frac{n}{2}\gamma^{latt}(\beta)
+ \left(\frac{1}{2}\gamma^{latt}(\beta)+\beta W^{latt}(\beta)\right)
h \frac{\partial}{\partial h}
\bigg]
\Gamma_{latt}^{(n)}\,,
\end{multline}
and analogously for the $\overline{MS}$-theory  by using
$W^{\overline{MS}}(\beta)$  and $\gamma^{\overline{MS}}(\beta)$ (in
order to avoid confusion with
the coupling costant, and in agreement with the literature on the subject, we
denote the $\beta$-function as $W(\beta)$).
By using the condition (\ref{rino}), we are able to join together the
$\beta$ and $\gamma$-function on the lattice to those in
$\overline{MS}$-scheme. Indeed we find
\begin{align}
\label{betas}
W^{\overline{MS}}(Z_{1}^{-1}\beta)
&=
\left(  Z_{1}+
\frac{1}{\beta}\frac{\partial Z_{1}}{\partial \beta^{-1}}\right)
W^{latt}(\beta)
\\
\gamma^{\overline{MS}}(Z_{1}^{-1}\beta)
&=\gamma^{latt}(\beta)-
\frac{1}{Z_{2}}
\frac{\partial Z_{2}}{\partial \beta^{-1}}
W^{latt}(\beta)
\end{align}
The first of them is the important relation that allows us to express
the coefficients of the $\beta$-function on the lattice in terms of the
coefficients of the continuum theory.

Given the $\beta$-function for the  non-linear $\sigma$-model with $N$
the number of vector components, we expand it in power of the coupling
costant $1/\beta$ in a generic \emph{scheme} of regularization
\be\label{betaexp}
W^{\emph{scheme}}(\beta)
=-\frac{w_{0}}{\beta^2}-\frac{w_{1}}{\beta^3}
-\frac{w_{2}^{\emph{scheme}}}{\beta^4}+O(\beta^{-5})\;;
\ee
the first two coefficients have not the superscript $\emph{scheme}$
because they are universal,
they come from the calculation respectively  at
one and two loops (the term from order zero vanishes in two dimensions);
explicitly they are given by
\begin{align}
w_{0} & =\frac{N-2}{2\pi}\;,
&
w_{1} & =\frac{N-2}{(2\pi)^2}\;;
\end{align}
all the other terms are scheme-dependent; the $w_{n}^{\emph{scheme}}$
coefficient is associated with  $1/\beta^{n+2}$ term of series
expansion and  correspond to a computation at $(n+1)$ loops.  We
report here the known results in $\overline{MS}$-scheme
(see \cite{brezin3loop}, or \cite{3loop, 4loop} for other
references)
\be
w_{2}^{\overline{MS}}  =\frac{1}{4}\frac{N^2-4}{(2\pi)^3}.
\ee
We also expand in $1/\beta$ the two renormalization constants
\begin{align}
Z_{1}= &
Z_{1}^{(0)}+\frac{Z_{1}^{(1)}}{\beta}
+\frac{Z_{1}^{(2)}}{\beta^2}+O(\beta^{-3})\\
Z_{2}= & Z_{2}^{(0)}+\frac{Z_{2}^{(1)}}{\beta}
+\frac{Z_{2}^{(2)}}{\beta^2}+O(\beta^{-3})
\end{align}
With the above conventions on the series expansions, now we look at
(\ref{betas}) and we rewrite it as:
\be
\label{betaslatt}
W^{latt}(\beta)
=\frac{W^{\overline{MS}}(Z_{1}^{-1}\beta)}
{Z_{1}+\frac{1}{\beta}\frac{\partial Z_{1}}{\partial \beta^{-1}}}\,;
\ee
from this equation it can be seen that the coefficient of order $n$ of
the expansion of $W^{latt}$ (i.e. $w_{n-2}^{latt}$) can be evaluated
as long as one knows the coefficients of 
$W^{\overline{MS}}$ up the same order
(i.e. $w_{1}^{\overline{MS}}$, $w_{2}^{\overline{MS}}$, \ldots,
$w_{n-2}^{\overline{MS}}$) and performs the
computation on the lattice of the constants $Z_{1}$ and $Z_{2}$
up order $n-1$.\footnote{To be precise, only the constant $Z_{1}$
  is required. The expansion for $Z_{2}$ however comes out as a side
  result of the computation.}
So we can argue the general result:
\be
w_{n-1}^{latt}=w_{(n-
\textrm{loop})}^{latt}=
F\left(\{w_{i}^{\overline{MS}}\}_{i=\{0,1,\cdots,n-1\}};
\{Z_{1}^{(j)} 
\}_{j=\{0,1,\cdots,n-1\}}\right)\,.
\ee
For example, for the first scheme-dependent coefficient 
$w_{2}^{latt}$, from (\ref{betaslatt}) we find
\be\label{w2latt}
w_{2}^{latt}
=w_{0}\left((Z_{1}^{(1)})^2-Z_{1}^{(2)}\right)
+w_{1}Z_{1}^{(1)}+w_{2}^{\overline{MS}}\;.
\ee

\section{Evaluation of the constants of renormalization}

In order to obtain the perturbative expansion of the constants
$Z_{1}$ and $Z_{2}$, we use relation (\ref{rino}) 
for the two-point function 1PI. We proceed as follows: we compute
$\Gamma_{latt}^{(2)}$ at $n-1$ loops and, from the knowledge
of $\Gamma_{\overline{MS}}^{(2)}$ at the same order, 
and the requirement of validity of
(\ref{rino}),
we find 
$Z_{1}$ and $Z_{2}$ at $n-1$ loops.

For the continuum theory we consider the expansion 
\be
\Gamma_{\overline{MS}}^{(2)}(p, \beta, h ; \mu)=-\beta (p^2+h) +
\Pi_{\overline{MS}}^{(0)}(p,  h ; \mu)
+\frac{\Pi_{\overline{MS}}^{(1)}(p,  h ; \mu)}{\beta}+\dots\;;
\ee
we report the already known two-loop results \cite{3loop, 4loop} in
the case of $N=-1$
\begin{subequations}
\label{contipai}
\begin{align}
\Pi_{\overline{MS}}^{(0)}(p,  h ; \mu)=& \frac{1}{4\pi}(p^2-h)\log\frac{h}{\mu^2}\\
\begin{split}
\Pi_{\overline{MS}}^{(1)}(p,  h ; \mu)=&
\frac{1}{16\,\pi^2}\left(\log^2\frac{h}{\mu^2}
+8\log\frac{h}{\mu^2}-3+12\,(2\pi)^2 R\right)\,p^2\\  
& \quad - \frac{1}{8\pi^2}\left(\log^2\frac{h}{\mu^2} + \log\frac{h}{\mu^2}\right)\,h 
\end{split}
\end{align}
\end{subequations}
where $R$ is an integral defined as
\be
\label{erre}
\begin{split}
R &:= \lim_{h \to 0} \;h
\int_{-\infty}^{\infty} \frac{d p_{x}}{2  \pi} \int_{-\infty}^{\infty}
\frac{d p_{y}}{2  \pi}
\int_{-\infty}^{\infty} \frac{d q_{x}}{2  \pi} \int_{-\infty}^{\infty}
\frac{d q_{y}}{2  \pi}
\frac{1}{(p^2+h)(q^2+h)( (p+q)^2+h)}
\\
& \phantom{:}=\frac{1}{24 \, \pi^{2}} 
\psi^{\prime}\!\left(\smfrac{1}{3}\right)
 - \frac{1}{36}\;, 
\end{split}
\ee
with $\psi(z) = d \log \Gamma(z) / dz$, but it appears only in
intermediate stages of the computation and cancels out in any of the
results.
Therefore
\be
\label{eq.robaccia}
\begin{split}
&\lefteqn{Z_{2}\, \Gamma_{\overline{MS}}^{(2)}(p, Z_{1}^{-1}\beta,
  Z_{1}Z_{2}^{-1/2}h ; \mu)}\\
&= -\beta\, (p^2+h) + \frac{1}{4\pi}(p^2-h)\log\frac{h}{\mu^2} +
  \left(Z_{1}^{(1)}-Z_{2}^{(1)}\right) p^{2} -\frac{1}{2} Z_{2}^{(1)}
  h
\\
& \quad + \frac{1}{\beta} \bigg[ \frac{1}{16\,\pi^2}
 \bigg(\log^2\frac{h}{\mu^2}
+8\log\frac{h}{\mu^2}-3+12\,(2\pi)^2 R \bigg)\,p^2  -
  \frac{1}{8\pi^2}
  \bigg( \log^2\frac{h}{\mu^2} +
  \log\frac{h}{\mu^2} \bigg) \,h
\\
& \hphantom{\quad + \frac{1}{\beta} \left[\right.}
+ \bigg( Z_{1}^{(2)} -
  Z_{2}^{(2)} + Z_{1}^{(1)}Z_{2}^{(1)} 
- \left(Z_{1}^{(1)}\right)^{2}
  + 
\frac{Z_{2}^{(1)}}{4 \pi}\log\frac{h}{\mu^2}
 + \frac{Z_{1}^{(1)}}{4 \pi} - \frac{Z_{2}^{(1)}}{8 \pi} \bigg) 
\, p^{2}
\\
& \hphantom{\quad + \frac{1}{\beta} \left[\right.}
+ \bigg( \frac{1}{8} \left( Z_{2}^{(1)} \right)^{2} - \frac{1}{2}
  Z_{2}^{(2)} -  \frac{Z_{1}^{(1)}}{4 \pi}\log\frac{h}{\mu^2}  -
  \frac{Z_{2}^{(1)}}{8 \pi}\log\frac{h}{\mu^2} 
- \frac{Z_{1}^{(1)}}{4 \pi} + \frac{Z_{2}^{(1)}}{8 \pi}\bigg) \, h
\bigg].
\end{split}
\ee


\section{The triangular lattice}

\begin{figure}
\label{figTriangLatt}
\begin{center}
\setlength{\unitlength}{24pt}
\begin{picture}(16,4.5)
\put(0,0){\includegraphics[scale=0.6]{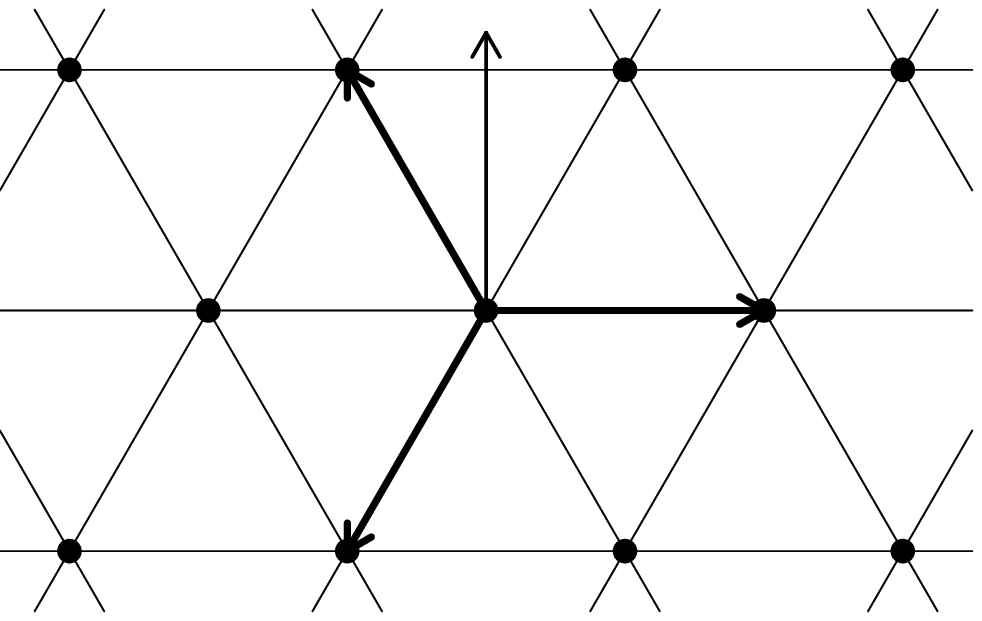}}
\put(3.95,1.9){$\e_1\equiv\hat{\e}_x$}
\put(2.45,3){$\e_2$}
\put(2.45,1.5){$\e_3$}
\put(3.05,3.3){$\hat{\e}_y$}

\put(9,0){\includegraphics[scale=0.6]{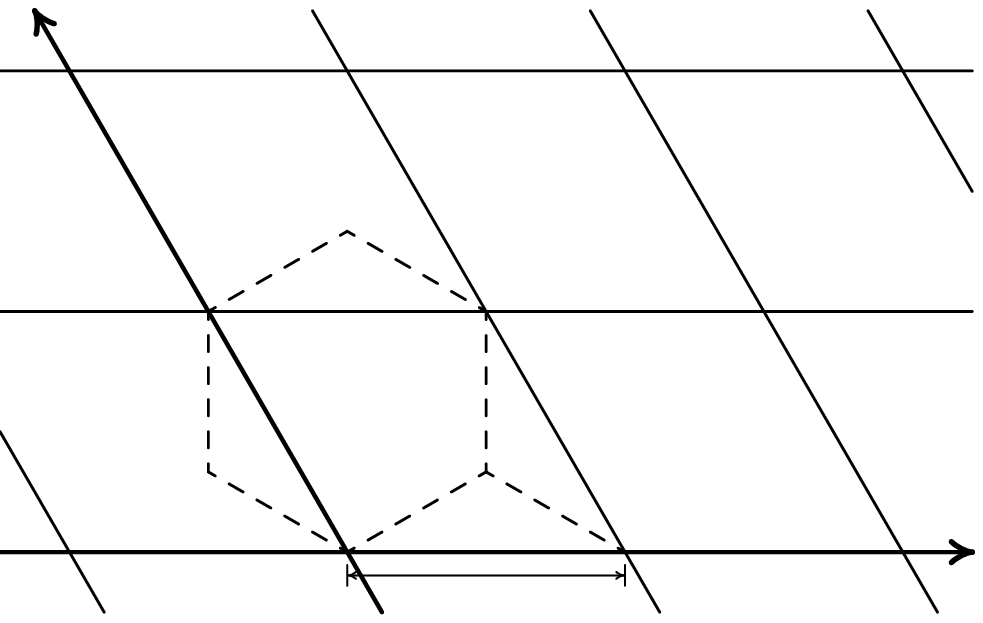}}
\put(15.7,0.9){$k_1$}
\put(9.6,4.3){$k_2$}
\put(12.4,0.1){$\scriptstyle{2 \pi}$}
\put(12.75,1.3){$\scriptstyle{\flat}$}
\put(10.75,1.3){$\scriptstyle{\flat}$}
\put(12.5,0.8){$\scriptstyle{\sharp}$}
\put(11.5,2.532){$\scriptstyle{\sharp}$}
\end{picture}
\caption{\small Left: the cartesian basis and the redundant basis on
  the triangular lattice. Right: Brillouin zone in momentum space. The
  rhombus or the hexagon are equivalent choices, as the pairs of
  triangles denoted with $\flat$ and $\sharp$ are related resp.~by
  periodicity in $k_1$ and $k_2$. While the hexagon corresponds to the
  direct construction of the reciprocal lattice, the rhombus is
  computationally convenient, as it is a product of one-dimensional
  intervals.
\label{basis}
}
\end{center}
\end{figure}

On a triangular lattice each site has 6 neighbours. It is convenient
to introduce a redundant basis of three vectors $\e_{(i)}$, as shown
in figure~\ref{basis}, such that $\sum_{i} \e_{(i)} = \boldsymbol{0}$,
$\e_{i}\cdot\e_{i}=1$, and if $i\neq j$ then $\e_{i}\cdot\e_{j}=
-\frac{1}{2}$.

Lattice sites are labelled by three integers $\{n_{i}\}$, with 
$\bx = \sum_{i} n_{i} \e_{(i)}$.
Because of redundancy, a constant can be added to the $n_{i}$'s
without changing $\bx$, i.e. there is an equivalence relation
\be
\left(n_{1}, n_{2}, n_{3}\right) \sim \left(n_{1}+m, n_{2}+m, n_{3}+m\right).
\ee
A representative of each class is chosen, for example, by fixing $n_{3}=0$, as
\be
\left(n_{1}, n_{2}, n_{3}\right) \sim \left(n_{1}-n_{3}, n_{2}-n_{3},
0\right). 
\label{gauge}
\ee
Remark that
\be
\bx \cdot \bx = \frac{3}{2} \bigg[ \sum_{i} n_{i}^{2} -
  \frac{1}{3} \Big( \sum_{i} n_{i} \Big)^{2} \bigg] .
\ee
Similarly, the conjugate quantity $\bk = \frac{2}{3}\sum_{i} k_{i}
\e_{(i)}$ is characterized by the three numbers $k_{i}$, such that 
$\sum_{i} k_{i} =0$. The factor $\frac{2}{3}$ is introduced to have
\be
\bk \cdot \bx = \sum_{i} k_{i} n_{i}.
\ee
As a consequence $|k_{i}|<\pi$ and 
\be
\bk \cdot \bk = \frac{2}{3} \sum_{i} k_{i}^{2}
\ee
so that the domain for $\bk$ is a hexagon of side $2 \pi/\sqrt{3}$, or
equivalently a $(\pi/6)$-angle
rhombus of sides $2 \pi$ (cfr.\ figure~\ref{basis}).

Now we can introduce the Fourier  transform $\tilde{f}(\bk)$ for a function $f$ on the triangular lattice
\be
\tilde{f}(\bk) = \sum_{\rm{sites}} e^{-i \bk\cdot\bx} f(\bx)
\ee
which is such that
\be
f(\bx) = \frac{\int_{\rm{hexagon}} d^{2}\bk \,e^{i \bk\cdot\bx} \tilde{f}(\bk)}{\int_{\rm{hexagon}} d^{2}\bk}
\ee
By specializing this general formula to the gauge~\reff{gauge} we get
\be
f(\bx) = \int_{-\pi}^{\pi} \frac{d k_{1}}{2  \pi} \int_{-\pi}^{\pi} \frac{d k_{2}}{2  \pi}\, e^{i [k_{1} (n_{1}-n_{3})+ k_{2}(n_{2}-n_{3})]}
\tilde{f}(k_{1},k_{2},-k_{1}-k_{2})
\ee
where we substituted $k_{3}=-k_{1}-k_{2}$ and we kept into account the angle of $2 \pi/3$ between the vectors $\e_{(1)}$ and $\e_{(2)}$ in the integration
measure. Remark that the volume of the elementary cell generated by $\e_{(1)}$ and $\e_{(2)}$ will pop out once more in the continuum limit, indeed
\begin{eqnarray}
 \sum_{\rm{sites}} &\to & \frac{2}{\sqrt{3}}\,\int d^{2} \bx \\
  \int_{-\pi}^{\pi} \frac{d k_{1}}{2  \pi} \int_{-\pi}^{\pi} \frac{d k_{2}}{2  \pi} & \to &
    \int_{-\infty}^{\infty} \frac{d k_{1}}{2  \pi} \int_{-\infty}^{\infty} \frac{d k_{2}}{2  \pi} =
   \frac{\sqrt{3}}{2}\,    \int_{-\infty}^{\infty} \frac{d k_{x}}{2  \pi} \int_{-\infty}^{\infty} \frac{d k_{y}}{2  \pi}\label{kc}
\end{eqnarray}

\section{Tree level}
\def\psibar{{\bar{\psi}}}

To compare with~\reff{model} let us change the normalization of the
Grassmann fields to get for the free part of the action on the
triangular lattice 
\be
-\sum_{\rm{sites}} \beta_{t} \,\left\{ \sum_{i} \psibar(\bx) \left[ 2
  \psi(\bx)  - \psi(\bx + \e_{i}) - \psi(\bx - \e_{i})\right] + h_{t}
\psibar(\bx) \psi(\bx) \right\} 
\ee
which becomes by Fourier transform
\be
-  \int_{-\pi}^{\pi} \frac{d k_{1}}{2  \pi} \int_{-\pi}^{\pi} \frac{d
  k_{2}}{2  \pi} \beta_{t} \, \psibar(\bk) \left[\widehat{k}^{2}+
  h_{t}\right] \psi(\bk)
\ee
where
\be
\widehat{k}^{2} := \sum_{i} \widehat{k}_{i}^{2}:= \sum_{i} \left[2
  \sin\left(\frac{k_{i}}{2}\right)\right]^{2} = \sum_{i} \left(2-2\cos
k_{i}\right)
\ee
By using $\widehat{k}^{2} \approx \frac{3}{2} \bk^{2}$ and \reff{kc}
this  becomes in the continuum limit
\be
 - \frac{2}{\sqrt{3}}\,    \int_{-\infty}^{\infty} \frac{d k_{x}}{2
   \pi} \int_{-\infty}^{\infty} \frac{d k_{y}}{2  \pi} \,\beta_{t} \,
 \psibar(\bk) \left[
\frac{3}{2} \bk^{2}+ h_{t}\right] \psi(\bk)
 \ee
and it must be compared with the continuos expression
\be
- \int_{-\infty}^{\infty} \frac{d k_{x}}{2  \pi}
\int_{-\infty}^{\infty} \frac{d k_{y}}{2  \pi} \,\beta\,  \psibar(\bk)
\left[\bk^{2}+ h\right] \psi(\bk)
\ee
from which
we get the identifications (see also~\cite{oldmine})
\begin{align}
\beta_{t} &\equiv \frac{\beta}{\sqrt{3}}
&
h_{t} &\equiv \frac{3}{2} h 
\label{ht}
\end{align}
In the following it will be useful the evaluation of the integral
\be
I(h_{t}) :=  \int_{-\pi}^{\pi} \frac{d p_{1}}{2  \pi}
\int_{-\pi}^{\pi} \frac{d p_{2}}{2  \pi} \,\frac{1}{\widehat{p}^{2}+
  h_{t}}
\ee
in the limit of small $h_{t}$.
Using the relation 
\be
\cos p_{1} +\cos p_{2}=2\cos \frac{p_{1}+p_{2}}{2}\cos \frac{p_{1}-p_{2}}{2} 
\ee 
we rewrite the denominator
\be
\phat^2 +h_{t} = 6-4\cos \frac{p_{1}+p_{2}}{2}\cos
\frac{p_{1}-p_{2}}{2}-2\cos (p_{1}+p_{2}) + h_{t}
\ee
and then we make the change of variables
$k_{1}=\frac{p_{1}+p_{2}}{2}$ and $k_{2}=\frac{p_{1}-p_{2}}{2}$; 
the Jacobian of the transformation is $2$, but it simplifies with the
factor $1/2$ coming
from the new area of integration; in fact $k_{1}$ and $k_{2}$ run
inside the rhombus of vertices $(\pi,0),(0,\pi),(-\pi,0),(0,-\pi)$,
the Brillouin zone, which is contained twice in 
the square area $[-\pi,\pi]^2$. So we obtain
\be 
I(h_{t}) =\int_{-\pi}^{\pi} \,\frac{d k_{1}}{2  \pi} \int_{-\pi}^{\pi}
\frac{d k_{2}}{2  \pi}
\frac{1}{6-4\cos k_{1} \cos k_{2} -2\cos (2k_1) +h_{t}} \quad .  
\ee
We are now able to integrate in $k_{2}$ using the result 
\be
\int_{-\pi}^{\pi} \frac{d \theta}{2 \pi} 
\frac{1}{\alpha + \beta \cos \theta}= \frac{1}{\sqrt{\alpha^2 -\beta^2}}\quad , 
\ee 
we have
\begin{eqnarray}
I(h_{t}) & = &   \int_{-\pi}^{\pi} \frac{d k_{1}}{2 \pi}
\frac{1}{2\sqrt{(3-\cos (2k_{1})+\frac{h_{t}}{2})^2-4\cos^2 k_{1} }}\\
\nonumber & = &   \int_{-\pi}^{\pi} \frac{d k_{1}}{2 \pi}
\frac{1}{2\sqrt{(\frac{h_{t}+6}{2}+2\sin^2 k_{1} )^2 - (h_{t}+9) }}\\ \nonumber
& = &   \int_{0}^{2\pi} \frac{d k_{1}}{2 \pi}
\frac{1}{2\sqrt{(\frac{h_{t}+8}{2}-\cos k_{1})^2 - (h_{t}+9) }}
\end{eqnarray}
Finally, after the change $\cos k_{1}=x$, we can express our integral
by an elliptic integral~\footnote{From 3.148.2 of~\protect\cite{Table}  \be \int_{d}^{u}
dx \frac{1}{\sqrt{(a-x)(b-x)(c-x)(x-d)}}=\frac{2}{\sqrt{(a-c)(b-d)}}
F(\beta,r) \ee with $a>b>c\ge u>d $ and $\beta=\arcsin
\sqrt{\frac{(a-c)(u-d)}{(c-d)(a-u)}}$
$r=\sqrt{\frac{(a-b)(c-d)}{(a-c)(b-d)}}$. In our case
$a=\frac{h_{t}+8}{2}+\sqrt{h_{t}+9},\quad b=\frac{h_{t}+8}{2}-\sqrt{h_{t}+9},\quad
c=u=1,\quad d=-1$.\\
$F(\beta,r)=\int_{0}^{\beta} \frac{d\theta}{\sqrt{1-r^2
\sin^2\theta}}$ is the\emph{ elliptic integral of the second kind},
and if $\beta=\frac{\pi}{2}$ , $ F(\frac{\pi}{2},r)=K(r)$ is called
the \emph{complete} integral.}
\begin{eqnarray}
I(h_{t}) & = & \frac{1}{2\pi}\int_{-1}^{1} dx
\frac{1}{\sqrt{(1-x^2)(\frac{h_{t}+8}{2}+\sqrt{h_{t}+9}-x)(\frac{h_{t}+8}{2}-\sqrt{h_{t}+9}-x)}}\\
\nonumber & = & \frac{1}{2\pi}
\frac{2}{\sqrt{6+2\sqrt{h_{t}+9}+3h_{t}+\frac{h_{t}^2}{4} }}
K\left(\sqrt{\frac{4\sqrt{h_{t}+9}}{6+2\sqrt{h_{t}+9}+3h_{t}+\frac{h_{t}^2}{4}
}}\right)
\end{eqnarray}

When $h_{t}\to 0$ 
\be 
I(h_{t})=-\frac{1}{4\sqrt{3}\,\pi} \log \left(\frac{h_{t}}{72}\right) + O(h_{t}\log h_{t})   
\ee
and therefore, because of~\reff{ht}
\be 
I(h_{t}) \approx -\frac{1}{4\sqrt{3}\,\pi} \log \left(\frac{h}{48}\right)
\ee
We will also need the evaluation of the integral
\be
I_{2}(h_{t}) :=  \int_{-\pi}^{\pi} \frac{d p_{1}}{2  \pi} \int_{-\pi}^{\pi} \frac{d p_{2}}{2  \pi} \,\frac{1}{\left(\widehat{p}^{2}+ h_{t}\right)^{2}}
\ee
in the limit of small $h_{t}$. Of course
\be
I_{2}(h_{t})  = - \frac{\partial}{\partial h_{t}} I(h_{t}) = \frac{1}{4\sqrt{3}\,\pi\, h_{t}}  + O(\log h_{t})   
\ee
and therefore
\be
\lim_{h_{t}\to 0} h_{t}\, I_{2}(h_{t}) = \frac{1}{4\sqrt{3}\,\pi}.
\ee
Of course the divergent part could be obtained by going to the continuum limit
\be
h_{t}\int_{-\pi}^{\pi} \frac{d p_{1}}{2  \pi} \int_{-\pi}^{\pi} \frac{d p_{2}}{2  \pi} \,\frac{1}{\left[\widehat{p}^{2}+ h_{t}\right]^{2}}
 \sim h \, \frac{3}{2}\frac{\sqrt{3}}{2} \int_{-\infty}^{\infty} \frac{d p_{x}}{2  \pi} \int_{-\infty}^{\infty} \frac{d p_{y}}{2  \pi} \,
\frac{1}{\frac{9}{4}\left[p^{2}+h\right]^{2}} 
 \sim \frac{1}{4\sqrt{3}\,\pi}
\ee
Analogously
\be
\begin{split}
&
\lim_{h_{t} \to 0} \;h_{t}
\int_{-\pi}^{\pi}
 \frac{d p_{1}}{2  \pi}
 \frac{d p_{2}}{2  \pi}
 \frac{d q_{1}}{2  \pi}
 \frac{d q_{2}}{2  \pi} 
\, \frac{1}{(\phat^2+h_{t})(\qhat^2+h_{t})(
  \widehat{p+q}^2+h_{t})}
\\
& \quad \sim
\lim_{h \to 0} \;h\,
\bigg(\frac{\sqrt{3}}{2}\bigg)^{2} \bigg(\frac{2}{3}\bigg)^{2} 
\int_{-\infty}^{\infty}
 \frac{d p_{x}}{2  \pi} 
\frac{d p_{y}}{2  \pi}
 \frac{d q_{x}}{2  \pi} 
 \frac{d q_{y}}{2  \pi}
\, \frac{1}{(p^2+h)(q^2+h)( (p+q)^2+h)}
= \frac{R}{3}
,
\end{split}
\ee
where $R$ was defined in (\ref{erre}).

\section{One-loop diagrams}

\begin{figure}
\begin{center}
\setlength{\unitlength}{90pt}
\begin{picture}(2.7,0.85)
\put(0,0){
\includegraphics[scale=1.5, bb=0 100 160 150, clip=true]{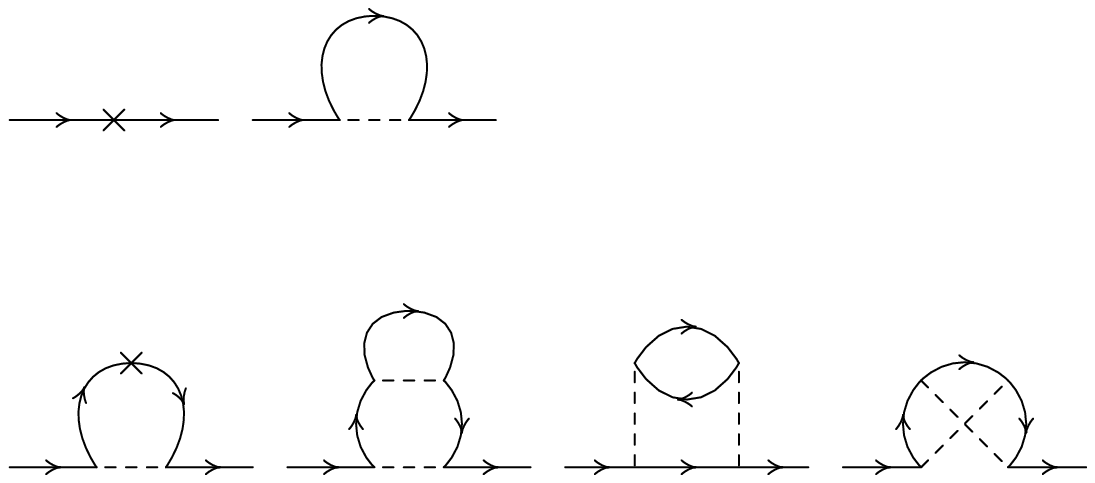}
}
\end{picture}
\caption{\small The Feynman diagrams for the two-point function at order 1.
\label{fig_feyndiagr1loop}
}
\end{center}
\end{figure}

The interaction terms on the triangular lattice are
\be
\int_{p} \psibar (\bp) \psi(\bp) - \frac{\beta_{t}}{2 } \int_{p,q,k}
\psibar(\bq+\bk)\psi(\bq) \,\widehat{k}^{2}\,
\psibar(\bp-\bk)\psi(\bp)
\ee
where we introduce the shorthand
\be
\int_{k} := \int_{-\pi}^{\pi} \,\frac{d k_{1}}{2  \pi}
\int_{-\pi}^{\pi} \frac{d k_{2}}{2  \pi}.
\ee
We wish to compute the 1PI two-point function. At one loop, two graphs
contribute (fig.~\ref{fig_feyndiagr1loop}).
On the triangular lattice, by defining
\be
\Delta(k) := \widehat{k}^{2} + h_{t}
\ee
we get
\be
\begin{split}
\Pi_{0}(p) 
&=
1 - \int_{k} \frac{\widehat{p+k}^{2}}{\Delta(k)}
=
1 - \int_{k}\frac{\widehat{p}^{2} + \widehat{k}^{2} - \frac{1}{2}
  \sum_{i}\widehat{p}_{i}^{2} \widehat{k}_{i}^{2} }{\Delta(k)}
\\
& =
1 - \widehat{p}^{2} I - 1 + h_{t} I + \frac{1}{6}
\widehat{p}^{2} \left( 1 - h_{t} I \right).
\end{split}
\ee
By going to the continuum limit, in the limit of small magnetic field
\begin{subequations}
\begin{eqnarray}
\Pi_{0}(p)  &\sim& \frac{\widehat{p}^{2}}{6} - \left[ \widehat{p}^{2}
  - h_{t}\right] \,I(h_{t}) \\
&\to & \frac{2}{\sqrt{3}} \left\{  \frac{3}{2}\,\frac{p^{2}}{6} +
\frac{3}{2} \left[p^{2} -h \right]\frac{1}{4\sqrt{3}\,\pi} \log\frac{h
  a^{2}}{48}\right\}\\
&=& \frac{p^{2}}{2\sqrt{3}} +  \frac{1}{4\pi}  \left[p^{2} -h
  \right]\log\frac{h a^{2}}{48}
\end{eqnarray}
\end{subequations}
By comparing the two expressions we obtain the one-loop result
\begin{eqnarray}
Z_{1} & = & 1 + \frac{3}{4 \pi \beta} \log \frac{\mu^{2}a^{2}}{48} +
\frac{1}{2 \sqrt{3}\, \beta} + 
{\cal O} \big( \smfrac{1}{\beta^{2}} \big) \\
Z_{2} & = & 1 + \frac{2}{4 \pi \beta} \log \frac{\mu^{2}a^{2}}{48} +
{\cal O} \big( \smfrac{1}{\beta^{2}} \big)
\end{eqnarray}
which, of course, result to be independent from the magnetic field.

\section{Two-loop diagrams}

\begin{figure}
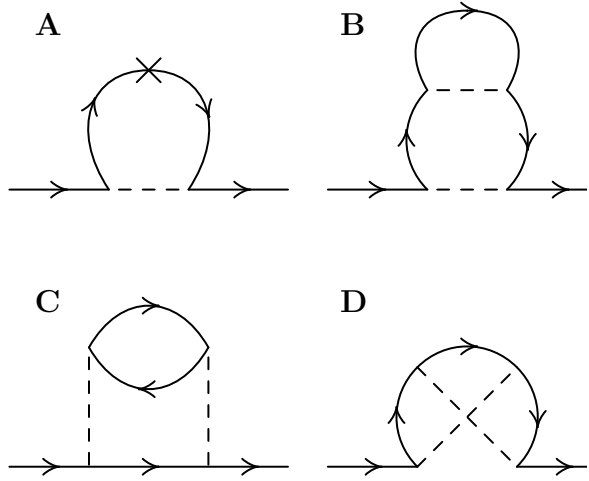

\begin{center}
\setlength{\unitlength}{105pt}
\begin{picture}(2.15,2)
\put(0,1){
\includegraphics[scale=1.5, bb=5 0 155 80, clip=true]{fig_feynman.eps}
}
\put(0,0){
\includegraphics[scale=1.5, bb=165 0 315 80, clip=true]{fig_feynman.eps}
}
\put(0.2,1.7){\bf A}
\put(1.3,1.7){\bf B}
\put(0.2,0.7){\bf C}
\put(1.3,0.7){\bf D}
\end{picture}
\caption{\small The Feynman diagrams for the two-point function at second
  order. On the top left corners, we report the identificative
  letters.
\label{fig_feyndiagr2loop}
}
\end{center}
\end{figure}

The diagrams at second order are the four ones shown in
figure~\ref{fig_feyndiagr2loop}.
As we expected these are the same Feynman diagrams that appear at the
second order of perturbative expansion of the
$\sigma$-model~\cite{3loop}.
According to the Feynman rules we
have to add a minus sign to the diagrams {\bf{A}} and {\bf{C}}: for
the first one since it has a mass insertion, for the second since it
has a loop.
So that the expression of the second order contribution of self-energy is
\be
\Pi_{1} = {\bf{A-B+C-D}}
\ee
with
\begin{align}
{\bf{A}}& =\int_{k}\frac{\widehat{p+k}^2}{\Delta(k)^2}\\
{\bf{B}}& =\int_{k,q}\frac{\widehat{p+k}^2
\widehat{k+q}^2}{\Delta(q)\Delta(k)^2}
\\
{\bf{C}}&
=\int_{k,q}\frac{(\widehat{q}^2)^2}{\Delta(p+q)\Delta(k)\Delta(k+q)}\\
{\bf{D}}& =\int_{k,q}\frac{\widehat{p+q}^2{\widehat{k}}^2}
{\Delta(q)\Delta(k-q)\Delta(p+k)}
\end{align}
The first two diagrams are easy to evaluate exactly in terms of $I$
and $I_{2}$. We find 
\begin{align}
{\bf{A}}
=  &\phat^2\left[-\frac{1}{6}I+I_{2}+\frac{1}{6}h_{t}I_{2}\right]
+I-h_{t}I_{2} \\
{\bf{B}} =  &
\phat^2\left[I^2-\frac{1}{2}I+I_{2}
+\frac{1}{3}h_{t}I_{2}-2h_{t}I_{2}I+\frac{1}{36}\right]+\\
& +2I-\frac{1}{6}-3h_{t}I^2-h_{t}I_{2}
+\frac{1}{2}h_{t}I-\frac{1}{6}h_{t}^2I_{2}+2h_{t}^2I_{2}I
\end{align}
The diagrams ${\bf{C}}$ and ${\bf{D}}$ are more involved.
First of all remark that
\begin{subequations}
\begin{align}
{\bf{D}} - {\bf{C}} 
& = \int_{k,q}\frac{\widehat{k}^2
  \big[\widehat{p+k+q}^2 - \widehat{k}^{2} \big] }
{\Delta(q)\Delta(k+q)\Delta(p+k)}
= \int_{k,q}\frac{\widehat{k}^2 \left[\Delta(p+k+q) - \Delta(k)\right]}
{\Delta(q)\Delta(k+q)\Delta(p+k)} \\
\label{eq.543654}
& \sim  \int_{k,q}\frac{\Delta(k) \left[\Delta(p+k+q) - \Delta(k)\right]}
{\Delta(q)\Delta(k+q)\Delta(p+k)} \\
\label{eq.543654b}
& = \int_{k,r}\frac{\Delta(k) \left[\Delta(p+r) - \Delta(k)\right]}
{\Delta(r-k)\Delta(r)\Delta(p+k)} 
\end{align}
\end{subequations}
where in (\ref{eq.543654}) we neglect terms of higher order in the
small-$h$ expansion.
We are interested in the first terms of the Taylor expansion for small
external momentum. We get
\be
\begin{split}
{\bf{D}} - {\bf{C}}
& \sim
\int_{k,r} \frac{1}{\Delta(r-k)} 
\bigg\{ 1 + \frac{4}{\Delta(k)}
\,\sum_{i} \sin p_{i} \sin k_{i} \sum_{j} \sin p_{j} \left[ \frac{\sin
    k_{j}}{\Delta(k)} -
 \frac{\sin r_{j}}{\Delta(r)}\right]  \vphantom{ \left(\sum_{i} \sin
  p_{i} \sin k_{i}\right)^{2}}
\\
 &
-
\frac{\Delta(k)}{\Delta(r)} +
\frac{\phat^{2}}{\Delta(r)} - \frac{1}{2\,\Delta(r)}\, \sum_{i}
\phat_{i}^{2}\widehat{k}_{i}^{2} -
\frac{4}{\Delta(k)\Delta(r)} 
\bigg(\sum_{i} \sin p_{i} \sin k_{i} \bigg)^{2} 
\bigg\}
\end{split}
\ee
We easily get
\begin{eqnarray}
\int_{k,r} \frac{1}{\Delta(r-k)} & = & I \\
- \int_{k,r} \frac{\Delta(k)}{\Delta(r-k) \Delta(r)} & = &   -2 I
+\frac{1}{6} + h_{t}I^2-\frac{h_{t}}{3}I\\
\int_{k,r} \frac{\phat^{2}}{\Delta(r-k) \Delta(r)} & = & \phat^{2}
I^{2} \\
- \frac{1}{2} \int_{k,r}  \frac{1}{\Delta(r-k) \Delta(r)}\, \sum_{i}
\phat_{i}^{2}\widehat{k}_{i}^{2} & = & \phat^{2}
\left(-\frac{I}{3}+\frac{1}{36}\right)
\end{eqnarray}
We have still to compute (changing $r$ into $-r$)
\be
4 \sum_{i,j} \sin p_{i} \sin p_{j} \int_{k,r}\frac{\sin
  k_{i}}{\Delta(r+k) \Delta(k)} \left[ \frac{\sin k_{j}}{\Delta(k)} +
  \frac{\sin r_{j}}{\Delta(r)} -
\frac{\sin k_{j}}{\Delta(r)}\right]  
\ee
The tensor form of the expression above is
\be
\sum_{i,j} \sin p_{i} \sin p_{j} \, \Lambda_{ij}
\ee
with $\Lambda_{ij}$ symmetric under the exchange of $i$ with
$j$, and permutation of indices $1,2,3$, so that
in general $\Lambda_{ij} = a + b \, \delta_{ij}$, which
substituted into the previous expression gives
$a (\sum_{i} \sin p_{i})^{2} 
+ b \sum_{i} \sin^{2} p_{i} \sim b \, \phat^{2} + O(p^{4})$
because we have that $\sum_{i } p_{i} =0$. Therefore we need only the
coefficient $b$ which can be computed, for example, as
\be
\Lambda_{11}-\Lambda_{13} = 4  \int_{k,r}\frac{\sin k_{1} - \sin
  k_{3}}{\Delta(r+k) \Delta(k)} \left[ \frac{\sin k_{1}}{\Delta(k)} +
  \frac{\sin r_{1}}{\Delta(r)} -
\frac{\sin k_{1}}{\Delta(r)}\right]  
\ee
Then we get
\begin{eqnarray*}
\int_{k,r} \frac{(\sin k_{1}-\sin k_{3}) \sin k_{1}}{\Delta(r+k)
\Delta^{2}(k) } & = & I \, \int_{k,r} \frac{(\sin k_{1}-\sin k_{3})
\sin k_{1}}{\Delta^{2}(k) }\\ 
& = & I \,\left[ \frac{1}{2} (I - h_{t
}I_{2}) - \frac{1}{12} + \frac{1}{8 \sqrt{3}\,\pi}\right]
\end{eqnarray*}
and
\begin{eqnarray}
\int_{k,r} \frac{\sin^{2} k_{1}}{\Delta(r+k) \Delta(k) \Delta(r)} & =
& \frac{ I^{2}}{3} - I \left(\frac{1}{6} - \frac{1}{2\sqrt{3}\,\pi}
\right) - \frac{R}{9} - \frac{G}{4} \\ 
\int_{k,r} \frac{\sin k_{1} \sin k_{3}}{\Delta(r+k) \Delta(k)
\Delta(r)} & = & - \frac{I^{2}}{6} + \frac{I}{4\sqrt{3}\,\pi} +
\frac{R}{18} - \frac{G}{8} - \frac{ 1 }{144}\\ 
\int_{k,r} \frac{\sin k_{1} \sin r_{1}}{\Delta(r+k) \Delta(k)
\Delta(r)} & = & - \frac{ I^{2}}{6} + I \left(\frac{1}{12} -
\frac{1}{4\sqrt{3}\,\pi} \right) + \frac{R}{18} + \frac{G}{8} +
\frac{L}{24} \\ 
\int_{k,r} \frac{\sin k_{3} \sin r_{1}}{\Delta(r+k) \Delta(k)
\Delta(r)} & = & \frac{I^{2}}{12} -\frac{I}{8\sqrt{3}\,\pi} -
\frac{R}{36} + \frac{G}{16} - \frac{K}{16} - \frac{L}{48} +
\frac{1}{288}
\end{eqnarray}
with
\begin{eqnarray}
G & := & \int_{k,r} \frac{\widehat{ k_{1}+ r_{1} }^{4} \left[
    \Delta(r+k) -\Delta(k) -\Delta(r)\right]}{\Delta^{2}(r+k)
  \Delta(k) \Delta(r)}\\
K & := & \int_{k,r} \frac{\widehat{k_{1}}^{2}  \widehat{k_{2}}
  \widehat{k_{3}} \widehat{r_{1}} \widehat{k_{1}+r_{1} }}{\Delta(r+k)
  \Delta(k) \Delta(r)}\\
L & := & \int_{k,r} \frac{\widehat{k_{1}+r_{1}}^{2}
  \widehat{k_{1}}^{2} \widehat{r_{1}}^{2}}{\Delta(r+k) \Delta(k)
  \Delta(r)}
\end{eqnarray} 
So finally we found:
\be
\label{ci}
{\bf{D}} - {\bf{C}} =  \phat^{2} \left[ I \left(\frac{1}{3} -
  \frac{3}{2\sqrt{3}\,\pi} \right)  +  R + \frac{3\,G +K+L}{4}  
   - \frac{1}{72} \right]  - I +\frac{1}{6} +
h_{t}I^2-\frac{h_{t}}{3}I
\ee
and in conclusion
\be
\label{eq.atlast}
\begin{split}
\Pi_{1}
&=
\phat^{2} \left[I^{2}  - \frac{2\,I}{\sqrt{3}\,\pi}   +
  \frac{1}{72} + \frac{1}{24 \sqrt{3}\,\pi}+  R+ \frac{3\,G +K+L}{4}
  \right]\\
& \quad
+ h_{t} \left[ -2 I^{2} +I\left(\frac{1}{6}+ \frac{1}{2
    \sqrt{3}\,\pi} \right) - \frac{1}{24\sqrt{3}\,\pi}\right]
\end{split}
\ee
By comparing the two expressions (\ref{eq.robaccia}) and
(\ref{eq.atlast}) we obtain the two-loop result
\begin{align}
Z_{1}^{(2)} 
& = 
\frac{9 }{16 \pi^2 } \log ^2 \frac{\mu^{2}a^{2}}{48}+\frac{\sqrt{3}
}{4 \pi } \log \frac{\mu^{2}a^{2}}{48}+\frac{3}{8 \pi^2 }\log
\frac{\mu^{2}a^{2}}{48} +
\frac{3 \left( 3\,G +K+L\right)}{4}+\frac{3}{16 \pi ^2}
+\frac{1}{8}\\
Z_{2}^{(2)} 
& = 
\frac{5 }{16 \pi^2 } \log ^2 \frac{\mu^{2}a^{2}}{48}+\frac{1 }{4
  \sqrt{3}\,\pi } \log \frac{\mu^{2}a^{2}}{48}
\end{align}
and the three-loop result by \reff{w2latt}
\be
w_{2}^{latt} = \frac{1}{16 \pi }-\frac{\sqrt{3}}{8 \pi^2}+\frac{3}{16
  \pi ^3} + \frac {9 \left( 3\,G +K+L\right)}{8 \pi }.
\ee
By application of the coordinate-space method by L\"uscher and
Weisz~\cite{luscher} suitably modified for the triangular lattice (see
appendix~\ref{LW}) we
have obtained the numerical determinations
\begin{subequations}
\label{eq.numGKL}
\begin{eqnarray}
G & = & -0.025786368\\
K & = & -0.007632210\\
L & = & \hphantom{-}0.035410394
\end{eqnarray}
\end{subequations}
with errors smaller than the quoted digits,  from which we recover the value
\be
w_{2}^{latt} = -0.01375000819\, .
\ee

\section{A direct application}

The determination of the coefficient $w_{2}^{latt}$ can be used, as
shown in Ref.~\cite{alanpotts}, to recover, for example, the phase
boundary in the plane of complex temperature for the $q$-state Potts
model in the limit $q\to 0$. This separatrix is, indeed, a special
renormalization-group flow curve. If we call $x$ and $y$,
respectively, the real and imaginary part of the complex temperature
we must have therefore that
\be
y(x) = y_0 \, \left( 1 + \frac{A_1}{x} + \frac{A_2}{x^2} + \cdots
\right) \label{prima}
\ee
where
\begin{align}
A_1 = &\,  \frac{\frac{w_1}{3}}{ - \frac{w_0}{\sqrt{3}} } = -
\frac{1}{2 \, \pi\, \sqrt{3}}
&
A_2 = &\,  \frac{\frac{w_2^{latt}}{3 \sqrt{3}}}{ -
  \frac{w_0}{\sqrt{3}} } 
= 
\frac{2 \pi}{9} \,
w_{2}^{latt} 
\end{align}
and $y_0$ was numerically estimated to be
\be
y_0 \approx 0.394 \pm 0.004
\ee
For numerical purposes in~\cite{alanpotts} a variant parametrization
is followed, that is
\be
y(x) = y_0 \, \exp \left[ 1 + \frac{B_1}{x-\alpha_0} +
  \frac{B_2}{(x-\alpha_0)^2} + \frac{B_3}{(x-\alpha_0)^3} +\cdots
  \right]
\label{seconda}
\ee
where comparison with~\reff{prima} in the limit of large $x$ gives the
relations
\begin{align}
B_1 = &\,  A_1 = - \frac{1}{2 \, \pi\, \sqrt{3}}
&
B_2 = &\,  A_2  - \frac{A_1^2}{2} - \alpha_0\,A_1 \, .
\end{align}
The parameter $\alpha_0$, and $A_i$ and therefore $B_i$ with $i\geq 2$
were not known. In~\cite{alanpotts} the authors decided to
truncate~\reff{seconda} by setting $B_i =0$ for $i\geq 3$ and try
estimated $\alpha_0$ and $B_2$ by this ansatz by imposing   the value
of the function and its derivative on the last known numerical point,
that is $y(0.0198)= 0.23$ and $y'(0.0198)=0.369003$. They estimated 
\begin{align}
\alpha_0 \approx &\, -0.550842 \\
B_2 \approx &\, -0.122843 \, .
\end{align}
From our calculation we get an evaluation of $B_2$
\begin{align}
A_2 \approx &\, 0.00959932 \\
B_2 \approx &\, 0.0053776 + \,\frac{\alpha_0}{2 \, \pi\, \sqrt{3}}
\end{align}
so that we can use the strategy just discussed to derive $B_3$ in
addition to $\alpha_0$ and $B_2$.
We obtain 
\begin{figure}[ht!]
\begin{center}
\setlength{\unitlength}{27.5pt}
\begin{picture}(8.5,9.5)
\put(0,0){\includegraphics[scale=0.55, bb=91 98 515 576]{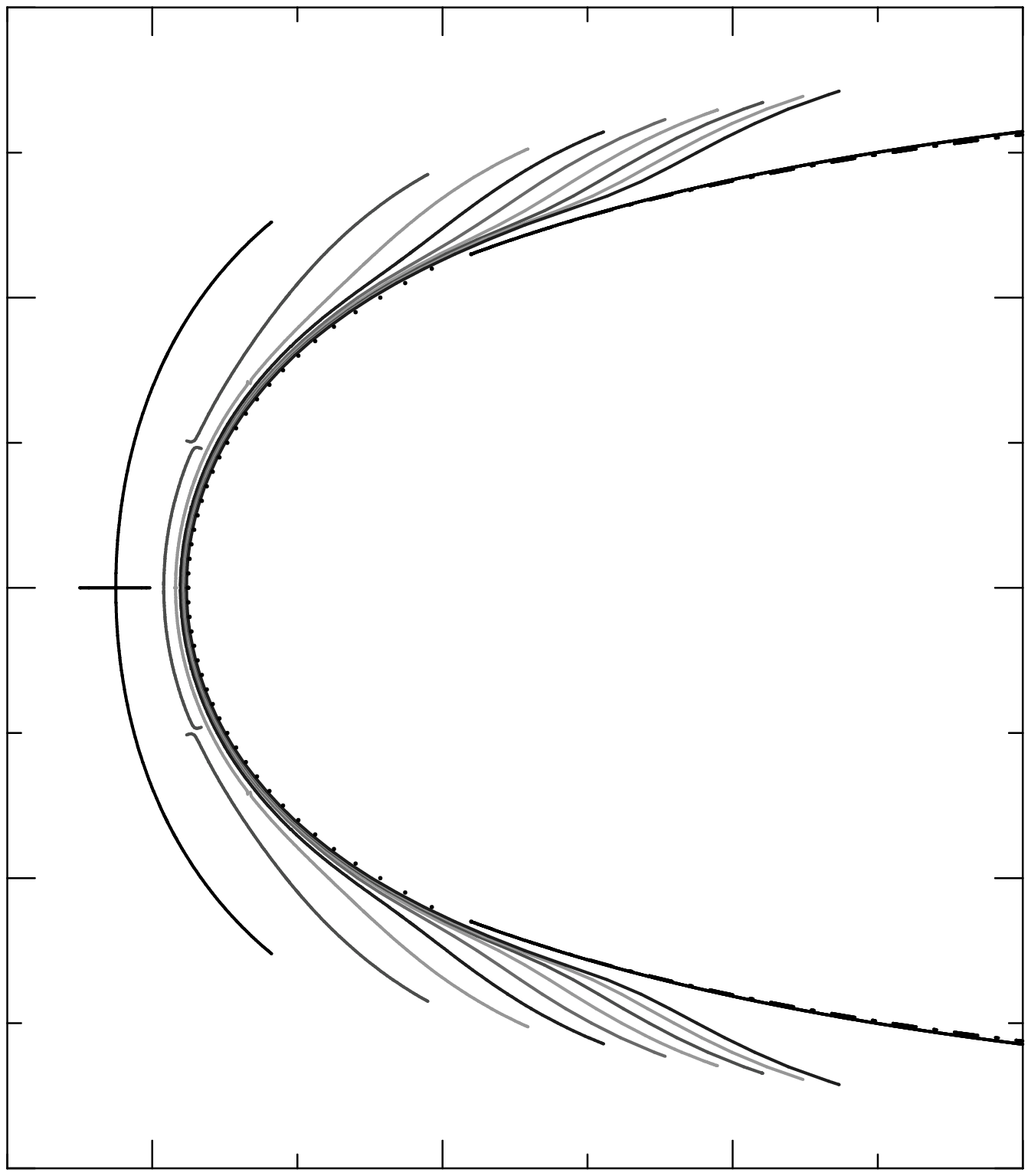}}
\put(1.45,0){\makebox[0pt][c]{$-0.2$}}
\put(3.75,0){\makebox[0pt][c]{$0$}}
\put(5.95,0){\makebox[0pt][c]{$0.2$}}
\put(8.15,0){\makebox[0pt][c]{$0.4$}}
\put(8.35,0.5){$\mathrm{Re}(1/t)$}
\put(0.2,0.3){\makebox[0pt][r]{$-0.4$}}
\put(0.2,2.5){\makebox[0pt][r]{$-0.2$}}
\put(0.2,4.7){\makebox[0pt][r]{$ 0  $}}
\put(0.2,6.9){\makebox[0pt][r]{$ 0.2$}}
\put(0.2,9.1){\makebox[0pt][r]{$ 0.4$}}
\put(0.1,8.3){\makebox[0pt][r]{$\mathrm{Im}(1/t)$}}
\end{picture}
\setlength{\unitlength}{40pt}
\begin{picture}(4,5)
\put(0.5,3.3){\includegraphics[scale=0.8, bb=121 13 325 146]{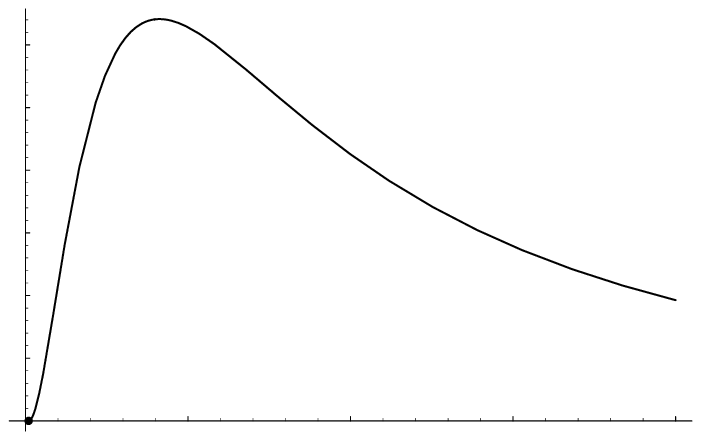}}
\put(0.60,3.2){$\scriptstyle{0}$}
\put(1.54,3.2){$\scriptstyle{1}$}
\put(2.48,3.2){$\scriptstyle{2}$}
\put(3.42,3.2){$\scriptstyle{3}$}
\put(4.36,3.2){$\scriptstyle{4}$}
\put(3.9,3.6){$\scriptstyle{\mathrm{Re}(1/t)}$}
\put(0.12,4.13){$\scriptstyle{0.001}$}
\put(0.12,4.85){$\scriptstyle{0.002}$}
\put(0.12,5.57){$\scriptstyle{0.003}$}
\put(0.55,5.93){$\scriptstyle{\delta}$}
\end{picture}
\end{center}
\caption{\small Phase boundaries for infinite strips of the triangular
  lattice. Numerical values from different lattice widths $L$, from 2
  to 9, on gray-tone curves from left to right.
   Black dots reproduce the extrapolated $L\to\infty$
  limiting curve in the
  region of negative $\hbox{Re\,} (1/t)$.
  The black dotted-dashed curve and the continuous black curve
  (almost indistinguishable), in the region
  of positive $\hbox{Re\,} (1/t)$,
  are respectively the old and new curves from the ansatz of
  equation~\reff{seconda}. In the magnification on the right, we plot
  the discrepancy $\delta$ between the two curves along the 
  $\hbox{Im\,} (1/t)$ axis, as a function of $\hbox{Re\,} (1/t)$: 
  as it should, it vanishes with its first
  derivative, at the numerical point $0.0198$ used for the extrapolation, and
  vanishes asymptotically because the same estimate of the
  asymptote $y_0$ is used; all in between, it remains of order $10^{-3}$.
\label{fig.rgt0}
}
\end{figure}
\begin{align}
\alpha_0 \approx &\, -0.778527 \\
B_2 \approx &\, -0.066160 \\
B_3 \approx &\, -0.162495 \, .
\end{align}
The curve resulting from this numerical values does not differ
substantially from the old one as can be seen in
Fig.~\ref{fig.rgt0}. This gives more confidence on the method and
results in~\cite{alanpotts}.

\section*{Acknowledgements}

We thank Jesus Salas and Alan Sokal for their interest in our work and
for providing us the numerical information needed to produce
figure~\ref{fig.rgt0}.

\appendix

\section{L\"uscher-Weisz method for evaluation of lattice integrals}
\label{LW}

In the evaluation of two-dimensional lattice integrals, we used the
coordinate method illustrated in the paper by L\"uscher and
Weisz~\cite{luscher}, and
specialized to two dimensions by Dong-Shin Shin \cite{shin1,shin2},
although also the momenta method proposed in the appendix~C
of~\cite{menotti} could have been used.

The main idea is the use of some basic relations for the free
propagator in coordinate space (the defining Laplacian equation and a
set of relations due to Vohwinkel), in order to find a recursion
which, starting from the values in a certain number of sites
neighbouring the origin (the {\em fundamental} lattice integrals),
allows to find the whole set of free propagators in lattice sites in a
large radius $R$, in a time which scales polynomially with $R$. As a
side result, it gives a simple proof of the fact that all these values
are linear combinations with rational coefficients of the fundamental
lattice integrals.

Generalization of the procedure to the triangular lattice is not
straightforward, and involves some delicate points. Some of them are:
\begin{itemize}
\item In the redundant set of variables $(p_1, p_2, p_3)$, the
  constraint $\sum_{i} p_{i}=0$ does not allow for derivatives in a
  single variable: one should either perform linear combinations of
  derivatives where the sum of coefficients is zero (for example,
  $(\nabla_1 - \nabla_3) f(p_1,p_2,p_3)$), or equivalently, perform
  derivation within a non-redundant choice of variables (for example,
  $\nabla_1 f(p_1,p_2,-p_1-p_2)$).
\item because of this fact, the Vohwinkel relations involve a larger
  number of terms, and thus it is more difficult to manipulate them in
  order to have a recursion relation. It will turn out that a larger
  strip is required for the first $x$-axis recursion.
\item For $G_{\ell}(x)$ at values of $\ell$ larger than 1, the choice
  of subtraction is now not anymore easily deduced by the Taylor
  expansion of the exponential and the requirement of periodicity. Now
  we also have the requirement of gauge-invariance under 
  $\bx \to \bx + m (1,1,1)$,
  which forces the application of ``hat'' factors only to
  combinations of $p_i$ where the sum of coefficients is zero.
\end{itemize}
Coming back to the point, the free subtracted propagator 
\be
G(x)=\int_{k}\,\frac{e^{i\boldsymbol{k}\cdot \boldsymbol{x}}-1}{\khat^2}
\ee
statisfies the Laplace equation
\be
\label{eq.lapl}
-\Delta G(x)= \delta^{(2)}(x)\;.
\ee
with the lattice operators
\be
\Delta :=  \sum_{i =1}^{3} \nabla^{*}_{i} \nabla^{\vphantom{*}}_{i}  =
\sum_{i =1}^{3} \left(\nabla^{\vphantom{*}}_{i} - \nabla^{*}_{i}
\right)
\ee
and we have
\be
G(0,0)=0
\qquad
G(1,0)=-\frac{1}{6}\, .
\ee
For the triangular function $H(x)$ defined as
\be
H(x)=  \int_{k}\,e^{i\boldsymbol{k}\cdot \boldsymbol{x}} \ln \khat^2
\ee
a set of Vohwinkel relations holds
\be
\label{eq.gght}
G(x+\widehat{\mu})-G(x-\widehat{i})-G(x+\widehat{\nu})+G(x-\widehat{\nu})=
(x_{\mu} - x_{\nu}) H(x)
\ee
but only two of them (e.g.~$(\mu,\nu)=(1,2)$ or $(1,3)$)
are independent.

Using the previous equations and the Laplace equation (\ref{eq.lapl}),
we are
able to eliminate $H(x)$, and write a recursion relation.  The one we
find on a width-2 strip along the $x$ axis is given by the set of
equations
\begin{align}
0 &=-6 \, G(x,1) + \sum_{\pm; \mu} G((x,1) \pm \hat{\mu}) \\
0 &=-6 \, G(x,0) + 2 (G(x+1,1) + G(x,1)) +
G(x + 1,0) + G(x - 1,0) 
\\
\begin{split}
0 &=(G(x-1,1) - G(x+1,1)) +x (G(x,2) - G(x,0))\\
& \qquad +(x-1) (G(x+1,2) - G(x-1,0))
\end{split}
\end{align}
which must be solved with respect to $G(x+1,a)$, with $a=0,1,2$, in
order to have a consistent recursion. A new fundamental integral is
required. A choice could be $G(2,1)$, which is valued
\be
G(2,1)= \frac{1}{3} - \frac{\sqrt{3}}{\pi}
\ee
In a similar fashion, given the values of $G(x)$ on the width-2 strip,
the function can be determined in the whole plane (a sector with
$x=(n,m)$, with $n\geq 2 m \geq 0$ is sufficient, because of
symmetry). The Laplacian equation alone is enough to fulfill this task. So we
conclude that at all values of $x$ the function $G(x)$ is in the set
$\mathbb{Q} + \frac{\sqrt{3}}{\pi} \mathbb{Q}$.

The integrals of the form
\be
K^{(2 n)}_1
=
\int_p
\frac{\phat_i^{2n}}{\phat^2}
\ee
which involve $G(x)$ only on the real axis, are easily computed, the
first values being
\begin{align}
K_1^4
&= 
-\frac{4}{3} + \frac{4 \sqrt{3}}{\pi}
&
K_1^6
&= 
16 - \frac{24 \sqrt{3}}{\pi}
&
K_1^8
&= 
-\frac{448}{3} + \frac{288 \sqrt{3}}{\pi}.
\end{align}
The next ingredient we need in order to calculate all the
triangular-lattice quantity arising from our diagrammatics is the
two-propagator function in coordinate space. It turns out that the
proper subtraction is the following
\be
G_2(x)=
\int_{k} 
\frac{e^{i\boldsymbol{k}\cdot \boldsymbol{x}}-1+\frac{1}{4}
\Big( (\khat^2 - 2 \khat_3^2) (x_1-x_2)^2 + \textrm{cyclics} \Big) }
{(\khat^2)^2}
\ee
The triangular-lattice Laplacian relation
still reads
\be
-\Delta G_2(x)= G(x)
\ee
while the Vohwinkel relation, still for
$(\mu,\nu)=(1,2)$ or $(1,3)$, is
\be
\label{eq.ggh2t}
G_2(x+\widehat{\mu})-G_2(x-\widehat{\mu})
-G_2(x+\widehat{\nu})+G_2(x-\widehat{\nu})=
-(x_{\mu} - x_{\nu})
\Big( G(x) + \frac{1}{4\sqrt{3}\, \pi} \Big)
\ee
(remark the presence of the corrective contribution $1/(4\sqrt{3}\, \pi)$
due to regularization).  At the aim of building the recursion, also in
this case it turns out that, as the ``support'' of the relations is
identical to the one of the triagular-lattice $G(x)$ case, the
independent lattice integrals still must be the ones located at the
points $x \in \{ (0,0), (1,0), (2,1) \}$.  The first two vanish
because of the subtraction, while the last one is computed
analytically, with the result
\be
G_2(2,1) = \frac{1}{4\sqrt{3}\, \pi}
\ee
and thus, as it is again a rational times $\sqrt{3}/\pi$, still the
function $G_2(x)$ at a generic point is in the set 
$\mathbb{Q} + \frac{\sqrt{3}}{\pi} \mathbb{Q}$.

The two-propagator analogues of the quantities $K^{2n}_1$
are the integrals of the form
\be
K^{(2 n)}_2
=
\int_p
\frac{\phat_i^{2n}}{(\phat^2)^2}
\ee
They still involve $G_2(x)$ 
only on the real axis, and thus are easily computed, the
first values being
\begin{align}
K_2^4
&= 
\frac{1}{3} - \frac{\sqrt{3}}{3 \pi}
&
K_2^6
&= 
-4 + \frac{8 \sqrt{3}}{\pi}
&
K_2^8
&= 
\frac{176}{3} - \frac{104 \sqrt{3}}{\pi}
\end{align}
We need also the lattice sums
\begin{align}
G 
& =
\sum_{x \in \mathbb{Z}^2} \left[\left(\nabla_{1}^{*} -
  \nabla_{1}^{\vphantom{*}}\right)^{2} G(x) \right] \left[ G(x)
  \right]^{2}\\
\begin{split}
K 
& =
2\,\sum_{x \in \mathbb{Z}^2} \left[\left(\nabla_{1}^{*} -
  \nabla_{1}^{\vphantom{*}}\right)\left(\nabla_{2}^{*} +
  \nabla_{2}^{\vphantom{*}}\right)\, G(x) \right]
\left[\left(\nabla_{3}^{*} + \nabla_{3}^{\vphantom{*}}
  \right)\,G(x)\right] \,G(x)
\\
&
\quad
+ \sum_{x \in \mathbb{Z}^2} \left[\left(\nabla_{1}^{*} -
  \nabla_{1}^{\vphantom{*}}\right)\left(\nabla_{2}^{*} +
  \nabla_{2}^{\vphantom{*}}\right)\,
\left(\nabla_{3}^{*} + \nabla_{3}^{\vphantom{*}}\right) \, G(x)
\right]\left[\,G(x)\right]^{2}
\end{split}
\\
L 
& =
\sum_{x \in \mathbb{Z}^2} \left[\left(\nabla_{1}^{*} -
  \nabla_{1}^{\vphantom{*}}\right) G(x) \right]^{3}
\end{align}
At this aim we need the full strength of coordinate method: we
evaluate the subtracted propagators $G(x)$ and $G_2(x)$ on lattice
points up to a given hexagon of side $r$ ($\sim 25$), exactly in terms
of rationals, in negligible computational time (${\cal O}(r^2)$), from
which we deduce the ${\cal O}(r^2)$ largest terms in the sums above,
while the remaining contribution is estimated from the large-distance
behaviour of the integrands. The numerical results are reported in
equations (\ref{eq.numGKL}).

\section{Details on two loop lattice integrals}

By using the identity
\be
2 \left[ \sin \alpha + \sin \beta + \sin \gamma \right] =
-\widehat{\alpha}\,\widehat{\beta}\,\widehat{\gamma} ,
\ee
valid when $\alpha + \beta + \gamma =0$, in the cases
$(\alpha,\beta,\gamma) = ( k_{1}, k_{2}, k_{3})$ or
$(k_{i},r_{i},-k_{i}-r_{i})$,
we easily get

\begin{gather*} 
4\,\int_{k,r} \frac{\left(\sum_{i} \sin k_{i}\right)^{2}} {\Delta(r+k)
\Delta(k) \Delta(r)} = \int_{k,r} \frac{\widehat{k_{1}}^{2}
\widehat{k_{2}}^{2}\widehat{k_{3}}^{2}}{\Delta(r+k) \Delta(k)
\Delta(r)} 
= 
- 2\, I \,\bigg(1 - \frac{2 \,\sqrt{3}}{\pi}\bigg) - 6
\,G - \frac{1}{6}\\
4\,\int_{k,r} \frac{\sum_{i,j} \sin k_{i} \sin r_{j} } {\Delta(r+k)
\Delta(k) \Delta(r)} = \int_{k,r} \frac{\widehat{k_{1}}
\widehat{r_{1}} \widehat{k_{2}} \widehat{r_{2}} \widehat{k_{3}}
\widehat{r_{3}} }{\Delta(r+k) \Delta(k) \Delta(r)} 
= 
I \,
\bigg(1 - \frac{2 \,\sqrt{3}}{\pi} \bigg) + 3\,G - \frac{3\,K}{2} +
\frac{1}{12}\\
4\,\int_{k,r} \frac{\left[\sin k_{1} +\sin r_{1} -\sin
\left(k_{1}+r_{1}\right)\right]^{2}} {\Delta(r+k) \Delta(k) \Delta(r)}
= \int_{k,r} \frac{\widehat{k_{1}+r_{1}}^{2} \widehat{k_{1}}^{2}
\widehat{r_{1}}^{2}}{\Delta(r+k) \Delta(k) \Delta(r)} 
= L
\\
4\,\int_{k,r} \frac{\prod_{i=1,2}\left[\sin k_{i} +\sin r_{i}- \sin
\left(k_{i}+r_{i}\right)\right]}{\Delta(r+k) \Delta(k) \Delta(r)} =
\int_{k,r} \frac{\widehat{k_{1}} \widehat{r_{1}} \widehat{k_{1}+r_{1}}
\widehat{k_{2}} \widehat{r_{2}} \widehat{k_{2}+ r_{2}} }{\Delta(r+k)
\Delta(k) \Delta(r)} 
= 
- \frac{L}{2} - \frac{3\,K}{2}
\end{gather*}
We also see that
\begin{eqnarray*}
- 2 \sum_{i}\left[ \sin k_{i} + \sin r_{i} - \sin
-\left(k_{i}+r_{i}\right)\right] &= &
-\sum_{i}\widehat{k_{i}}\widehat{r_{i}}\widehat{k_{i}+r_{i}}\\
& = &
\widehat{k_{1}} \widehat{k_{2}}\widehat{k_{3}} + \widehat{r_{1}}
\widehat{r_{2}}\widehat{r_{3}} - \widehat{k_{1}+r_{1}}
\widehat{k_{2}+r_{2}}\widehat{k_{3}+r_{3}}
\end{eqnarray*}
therefore
\be
\int_{k,r}
\frac{ \widehat{k_{1}} \widehat{r_{1}} \widehat{k_{1}+r_{1}}\sum_{i}
  \widehat{k_{i}} \widehat{r_{i}} \widehat{k_{i}+r_{i}}}{\Delta(r+k)
  \Delta(k)
\Delta(r)} = \int_{k,r}
\frac{-\widehat{k_{1}} \widehat{k_{2}} \widehat{k_{3}}\sum_{i}
  \widehat{k_{i}} \widehat{r_{i}} \widehat{k_{i}+r_{i}}
}{\Delta(r+k)
\Delta(k) \Delta(r)} = - 3 \,K
\ee
computed either using the first two lines or the last two lines of the
previous block of identities.



\begin{thebibliography}{99}

\bibitem{Temperley} 
H.\ N.\ V.\ Temperley, {\it Graph Theory and Applications,} 
(Ellis Horwood Series 1981).

\bibitem{Biggs}  
N. Biggs, {\em Algebraic Graph Theory,}
2nd ed.\ (Cambridge University Press,
1993).

\bibitem{Royle}
C.\ Godsil and G.\ Royle, {\it Algebraic Graph Theory,}
1st ed.\ (Springer Verlag,
2001).

\bibitem{Diestel}
R.\ Diestel, {\it Graph Theory,}
(Springer Verlag,
2000).

\bibitem{Potts}
  R.\ B.\ Potts,
  {\it Proc.\ Cambridge Phil.\ Soc.}\ {\bf 48}, 106 (1952).
  
\bibitem{Wu}
  F.\ Y.\ Wu,
  {\it Rev.\ Mod.\ Phys.}\ {\bf 54}, 235 (1982)
   and {\bf 55}, 315 (1983).
   
\bibitem{Wu2}
F.\ Y.\ Wu,
{\it J.\ Appl.\ Phys.}\ {\bf 55} (1984) 2421.

\bibitem{tutte} 
W.\ T.\ Tutte, 
 {\it Canad.\ J.\ Math.}\ {\bf 6},  (1953) 80.
 
 \bibitem{foata}
I.\ M.\ Gessel and B.\ E.\ Sagan,
{\it Electr.\ Jour.\ Comb.}\ {\bf 3} (1996) 2.

\bibitem{alantutte} 
A.\ D.\ Sokal,
 in {\em Surveys in Combinatorics, 2005}, ed.\ by B.\ S.\ Webb
(Cambridge University Press,
2005), 173,
[arXiv:math.CO/0503607].

\bibitem{Kirchhoff}
G.\ Kirchhoff,
{\it Ann.\ Phys.\ Chem.}\ {\bf 72}, 497 (1847).
    
\bibitem{Schrock}
  R.\ Shrock and F.\ Y.\ Wu,
  {\it J.\ Phys.\ A} {\bf 33}, 3881 (2000)
  [arXiv:cond-mat/0004341].
  
\bibitem{Glasser}
M.\ L.\ Glasser and F.\ Y.\ Wu,
{\it Ramanujian J.}~{\bf 10}, 205 (2005)
[arXiv:cond-mat/0309198].

\bibitem{noi} 
S.\ Caracciolo, J.\ L.\ Jacobsen, H.\ Saleur, A.\ D.\ Sokal and A.\ Sportiello,
{\it Phys.\ Rev.\ Lett.}\ {\bf 93} (2004) 080601 [arXiv:cond-mat/0403271].

\bibitem{Polyakov_75}  
  A.\ M.\ Polyakov,
  {\it Phys.\ Lett.}\ B {\bf 59}, 79 (1975).
    
 \bibitem{brezin}
  E.\ Br\'ezin, J.\ Zinn-Justin and J.\ C.\ Le Guillou,
  {\it Phys.\ Rev.\ D} {\bf 14}, 2615 (1976).
  
\bibitem{Brezin_76}  
  E.\ Br\'ezin and J.\ Zinn-Justin,
  {\it Phys.\ Rev.\ B} {\bf 14}, 3110 (1976).
  
\bibitem{Bardeen_76}
  W.\ A.\ Bardeen, B.\ W.\ Lee and R.\ E.\ Shrock,
  {\it Phys.\ Rev.\ D} {\bf 14}, 985 (1976).
  
\bibitem{saleur1}
J.\ L.\ Jacobsen and H.\ Saleur,
{\it Nucl.\ Phys.\ B}\ {\bf  716},  439 (2005) [arXiv:cond-mat/0502052].

\bibitem{saleur2}
J.\ L.\ Jacobsen and H.\ Saleur,
{\it Nucl.\ Phys.\ B}\ {\bf  743},  207 (2006) [arXiv:cond-mat/0512058].

\bibitem{saleur3}
Y.\ Ikhlef, J.\ L.\ Jacobsen and H.\ Saleur,
{\em A staggered six-vertex model with non-compact continuum limit},
[arXiv:cond-mat/0612037].
   
\bibitem{alanpotts}  
J.-L.\ Jacobsen, J.\ Salas and A.\ D.\ Sokal, 
{\it J.\ Stat.\ Phys.}\ {\bf 119}, (2005) 1153 [arXiv:cond-mat/0401026].
   
\bibitem{Chaiken}
S.\ Chaiken,
   {\it SIAM J.\ Alg.\ Disc.\ Meth.}\ {\bf 3}, 319 (1982).

\bibitem{Moon}
J.\ W.\ Moon,
   {\it Discrete Math.}\ {\bf 124}, 163 (1994).

\bibitem{Abdesselam}
A.\ Abdesselam, 
    {\it Adv.\ Appl.\ Math.}\ {\bf 33}, 51 (2004), [arXiv:math.CO/0306396].


\bibitem{hyperforests}
S.\ Caracciolo,  A.\ D.\ Sokal and A.\ Sportiello,
{\em Grassmann Integral Representation for Spanning Hyperforests},
[arXiv:0706.1509].

\bibitem{Duplantier_88}  
B.\ Duplantier and F.\ David,
   {\it J.\ Stat.\ Phys.}\ {\bf 51}, 327 (1988).

\bibitem{falctreves} M.\ Falcioni and A.\ Treves,
{\it Nucl.\ Phys.\ B} {\bf 265}, 671 (1986).

\bibitem{3loop} S.\ Caracciolo and A.\ Pelissetto,
{\it Nucl.\ Phys.\ B} {\bf 420} 141 (1994) [arXiv:hep-lat/9401015].

\bibitem{4loop} S.\ Caracciolo and A.\ Pelissetto,
{\it Nucl.\ Phys.\ B} {\bf 455} 619 (1995) [arXiv:hep-lat/9510015].

\bibitem{shin2}
Dong-Shin Shin,
{\it Nucl.\ Phys.\ B} {\bf 546} 669 (1999) [arXiv:hep-lat/9810025]  

\bibitem{4-loop_bis}  B.\ All\'es, S.\ Caracciolo, A.\ Pelissetto and M.\ Pepe,
{\it Nucl.\ Phys.\ B} {\bf 562}, 581 (1999) [arXiv:hep-lat/9906014].

\bibitem{brezin3loop}
E.\ Br\'ezin and S.\ Hikami, 
J.\ Phys.\ A {\bf 11} 1141 (1976).


\bibitem{oldmine} S.\ Caracciolo,
{\it Nucl.\ Phys.\ B} {\bf 180} 405 (1981).

\bibitem{Table} 
I.\ S.\ Gradshteyn, I.\ M.\ Ryzhik, \emph{Table of Integrals, Series, and products,}
7th ed.\ (Academic Press, 2007) 


\bibitem{luscher} M.\ L\"uscher and P.\ Weisz, 
{\it Nucl.\ Phys.\ B} {\bf 445} 429 (1995) [arXiv:hep-lat/9502017].

\bibitem{shin1}
Dong-Shin Shin,
{\it Nucl.\ Phys.\ B} {\bf 525} 457 (1998) [arXiv:hep-lat/9706014].

\bibitem{menotti} S.\ Caracciolo, A.\ Pelissetto and P.\ Menotti,
{\it Nucl.\ Phys.\ B} {\bf 375} 195 (1992).


\end{thebibliography}
\end{document}